\documentstyle[epsfig]{aipproc}

\begin{document}
\title{TAROT: A status report
}

\author{Michel Bo\"er, J.L. Atteia, M. Bringer, A. Klotz, 
C. Peignot$^*$,\\
R. Malina, P. Sanchez$^{\dagger}$,\\
H. Pedersen$^{\ddagger}$\\
G. Calvet, J. Eysseric, A. Leroy, M. Meissonier$^{\|}$\\
C. Pollas, J. de Freitas Pacheco$^{\P}$
}
\address{$^*$Centre d'Etude Spatiale des Rayonnements (CNRS)\\
BP 4346, F 31028 Toulouse Cedex 4, France\\
$^{\dagger}$Laboratoire d'Astronomie Spatiale (CNRS), Marseille, France\\
$^{\ddagger}$Copenhagen University Observatory, Denmark\\
$^{\|}$INSU-CNRS, Division Technique, Paris, France\\
$^{\P}$Observatoire de la C\^ote d'Azur, Nice, France
}

\maketitle

\begin{abstract}
TAROT-1 is an automatic, autonomous ground based observatory whose
primary goal is the rapid detection of the optical counterparts of cosmic
gamma-ray burst sources. It will be able to begin imaging any GRB
localization 8 seconds after receipt of an alert
from CGRO/BATSE or HETE-2. TAROT-1 will reach the 17th V magnitude in 10
seconds, at a 10$\sigma$ confidence level. TAROT will be able to observe
GRB positions given by Beppo-SAX or RXTE, EUV transients from ALEXIS
alerts, etc. TAROT will also study a wide range of secondary objectives and
will feature a complete automatic data analysis system, and a powerful
scheduling software. TAROT will be installed this fall on the Plateau du
Calern, 1200m above sea level. We report on the status of the
project.
\end{abstract}

\section*{Scientific goals}

\subsection*{Cosmic Gamma-Ray Bursts}

The T\'elescope \`a Action Rapide pour les Objets Transitoires (Rapid
Action Telescope for Transient Objects) has as primary objective the
detection of optical transients associated with gamma-ray bursts. Its
construction was decided in 1995, at a time when the optical emission of
GRBs was unknown, but predicted by a few models \cite{A}. 
At that time we thought that the best chance was to observe the
GRB source in optical during its gamma-ray activity, meaning a fast
moving wide field telescope. We also decided, given the various
constraints, to reach a sensitivity level somewhat below the theoretical
constraints, i.e. V$_{\rm{min}} \ge 16 {\rm\ or}\ 17$. The recent detection 
of the afterglow from GRB 970228 and GRB 970508 \cite{B,C} at optical and X-ray 
wavelengths demonstrates that this objective is achievable, since given the time 
delays involved, the light emitted by GRB 970508 would have been easily 
detectable by TAROT-1. Popular models \cite{D} invoke the chock of a 
relativistic fireball with the interstellar medium. TAROT will be able to detect 
it quite early, but also, it will detect the emission due to internal chocks 
within the fireball itself \cite{E}. In this later case, the optical emission is 
predicted to be simultaneous with the gamma-ray emission, and in the range of V 
magnitude between 14 and 16\cite{A}. With a maximum delay of 8 seconds from the 
burst 
onset to the TAROT observation, we will be able to catch 70\% of the sources 
while they are still active, provided that the error box is small enough (4 
square degrees). In the 
case of BATSE we hope to detect one or two bursts per year within this delay.

Detection of this emission by TAROT or by a similar experiment would be an 
important objective, since it may lead to the confirmation of the model, and 
give data on the physical source conditions. An early detection of a fireball at 
a relatively high level (e.g. magnitude 16) would trigger observations at  
larger telescopes, in order to take spectrum of the source itself, as well as of 
the 
host, and to detail the light curve of the optical transient from the 
beginning of the event, and to see the transition between the internal chock 
regime and the afterglow. 

Moreover, since TAROT has a large field of view (4 square degrees) and will 
operate 
continuously, it will be able to detect optical transients which may be related 
to undetected GRBs if the emission is beamed \cite{F}. Large areas of the sky 
will be surveyed both for secondary objectives and to establish a reference 
catalogue for later detection of GRB optical counterparts. This has two 
advantages. 1) In the case we detect an object within the error box of a GRB, 
from the inspection of our catalogue we can precise its nature, real new object 
or variable or flaring object active at the GRB time; 2) During this survey we 
will detect a large number of variable or new objects, since a substantial 
fraction of 
the sky will be observed every night. How we will be able to separate GRB 
afterglows from variable or flare stars is another problem which is currently 
under study, but the information will be in our data.

\subsection*{Secondary Objectives}

An automated, versatile telescope like TAROT has a wide range of possibles 
applications. 
Objects may be observed upon alert or in a systematic mode. In 
addition, the wide field of view will result in a lot of serendipitous 
detections.

In alert mode we shall try to identify EUVE transients detected by the ALEXIS 
satellite and so far of unknown nature, as well as X-ray transients upon alert 
from the SAX, 
RXTE, BATSE, HETE-2, INTEGRAL, MOXE, etc. 

In the routine mode we plan to observe systematically several late type flare 
stars 
in order to test our ability to detect optical transients. Our programme 
includes 
also the detection of supernovae, symbiotic stars, asteroids and comets. The 
detailed 
program of TAROT is currently being elaborated and will be made 
available through our server and in later publications.

\section*{Technical Design}

The actual design of TAROT is summarized in Table \ref{table1}. As it can be 
inferred from the table, the goal of observing GRBs while they are active has 
driven the design. We list below some technical features.

\begin{table}
\caption{Summary of TAROT technical data}
\label{table1}
\begin{tabular}{ll}
Aperture & 25cm, f/3.3\\
Field of view & 2 x 2 degrees\\
Optical resolution & 20 microns\\
Maximum time to slew to target & 3 seconds (180 degrees)\\
Maximum slew speed & 120 deg/s\\
Tracking speed & Adjustable $\alpha$ and $\delta$, range from 0 to 60 deg/s\\
Maximum acceptable wind speed & 80 km/h\\
Mount type & Equatorial\\
CCD size & 2048 x 2048, 3 x 3 cm\\
CCD pixel size & 15 x 15 microns\\
CCD readout noise & $\leq$ 10 e$^-$\\
CCD readout time (imaging mode) & 10 seconds\\
Filter wheel & 6 positions (B, V, R, I, B+V, R+I, Blank)\\
Limiting V magnitude & 17 @ 10$\sigma$ in 10s, 19 in 1 min.\\
Typical integration time (alert) & 20 seconds\\
Single exposure maximum integration time & 5 minutes\\

\end{tabular}
\end{table}

\begin{itemize}

\item Mechanics: The mechanical design has been extensively studied in order to 
ensure the stability and the reliability of the telescope. The requirement was 
that TAROT should be able to track (without a guiding star) an object for at 
least 5 minutes, without any noticeable displacement on the CCD camera. Also, 
TAROT will accommodate wind speeds as high as 80 km/h (50mph). The behaviour of 
the mount has been simulated to ensure that no vibrations are generated during 
the acceleration/deceleration phase.

\item Motors and controls: The drives have been chosen in order to accomodate
for the large accelerations needed by TAROT. They will be able to make a  
 move to any point in less than 3 seconds, meaning a maximum speed of 120 
degrees/second., and accelerations as large as 100 deg/s$^2$. For simplicity, we 
decided to use the same motors for the declination and right ascension axis. All 
drives, including the focus and filter wheel mechanisms are controlled from the 
telescope control software via a single PC card. An extensive protection of all 
electric and electronic parts is used against lightning.

\item Filter wheel: We use a custom designed filter wheel with 6 positions. In 
addition to a transparent position, a set of standard Cousins B,V, I filters 
will 
be used, and two wide band filters, of transmission approximating the overall 
band pass of the B+V and R+I filters.

\end{itemize}

\section*{Software}

The software is one of the most sensitive parts of TAROT, since it should run in 
complete autonomy. The interfaces for the alerts and routine 
observations will use the Web, the mail, and socket processes (for 
GCN/BACODINE). In 
addition, a local interface will be available, mainly for testing and debugging 
purposes. Our objective is that the telescope operates unmanned for periods as 
long as 3 months. Hence the control program will be responsible for night 
operations, day/night transition, calibrations, focusing, etc. This software 
will 
take into account the data from the environmental sensors to decide what 
operation to perform, and will run the telescope accordingly. Routine operations 
can be interrupted at any time to process an alert. In addition the control 
software will perform general tasks such as housekeeping, logging...

Routine observations, and follow-up alert observations will be scheduled through 
a particular software called the {\it Majordome}. This software 
implements several algorithms in order to ensure a maximum efficiency of the 
observations. Objects should be observed at minimal airmass (unless they are 
other constraints), and the number of possible observations should be maximized, 
according to various parameters such as the Moon, user constraints, observation 
types (periodic, repeated, time tagged...) and priorities. If an alerts occurs, 
the routine program is interrupted, and the alert processed according to a 
predefined sequence. The alert modifies in turn the input of the {\it Majordome} 
in order to introduce follow-up observations. 

We began to design a module to process automatically the data taken by TAROT. 
Our decision was based on the fact that TAROT will produce an average of 3Gb per 
night, and on the necessity to react quickly after an alert. This software will 
produce a list of sources detected in the image, together with their 
characteristics (photometry, spatial extension, apparent motion, etc.) will 
compare 
each object with the TAROT database (whenever possible), and with other 
available catalogues to search for a possible variability, or change in 
properties, or to detect candidate new objects. In order to ensure their nature, 
each observation will be done twice.

In addition to the above mentionned routine and alert mode, TAROT will be able 
to scan the sky according to two modes: in imaging mode we take a normal 2K x 2K 
image, and in scanning mode the telescope scans a wide area, while the CCD is 
read-out continuously. This later mode will be mainly used for BACODINE/BATSE 
alerts. In this mode a typical error box is scanned in less than 5 minutes. The 
scanning mode may be used also to build quickly a first database of TAROT 
objects, to a limiting magnitude of 17 (V).

\section*{Current Schedule}

The mechanics and the optics of TAROT have been delivered and integrated in 
September 1997 together with the drives. The software is currently being 
integrated and tested in the lab., and the optics will soon be submitted to 
interferometric measurements.

This fall (1997), the telescope will be moved to its final location, the 
"Plateau du Calern", 1200m above sea level and French Riviera. It will be 
installed in a building with a fully retractable roof, which has been recently 
refurbished in order to ensure maximum sky coverage.

After that the telescope will enter in an extensive test period (mechanics, 
software, security checks, optics, scientific validation...). During it we hope 
to be able to receive alerts at least through the GCN network. Routine 
scientific observations and automatic image processing should start running 
during the second semester of 1998.

\section*{Conclusion}

Though its dimensions are rather modest, TAROT will be a very efficient 
instrument, optimized for its prime objective, the detection of 
high energy transients. Given that 5 seconds are needed to obtain the coordinate 
information from BATSE/BACODINE or HETE-2,  TAROT will be able to get data from 
the source less than 8 second after the burst onset, while most sources are 
still active, and to eventually detect the internal chock from the GRB 
fireball. TAROT will be able also to estimate the background of transient events 
 over the sky, to detect putative "optical GRBs",  and to address a wide range 
of secondary objectives. Its schedule is well in accordance with  BATSE, SAX, 
RXTE and HETE-2 satellites.

{\it Acknowledgements}

The TAROT project is funded by the Centre National de la Recherche Scientifique 
(CNRS / INSU) in France, and by the Carlsberg Fondation in Denmark.

\end{document}